\documentclass[amsmath, amssymb,10pt, aps, prl, twocolumn, notitlepage, longbibliography, superscriptaddress]{revtex4-1}

\usepackage{graphicx}
\usepackage{calc}
\usepackage{braket}
\usepackage{slashed}
\usepackage{wasysym}
\usepackage{dsfont}
\usepackage{amsthm,amsmath,amsfonts,amssymb,verbatim,color}
\usepackage{lipsum}
\usepackage{bm}
\usepackage{epsfig,slashed}
\usepackage{hyperref}
\usepackage{flafter}

\newcommand{\bk}{{\bf k}}

\newcommand{\br}{{\bf r}}

\def\hbsigma{\hat{\boldsymbol \sigma}}

\newcommand{\hbk}{{\hat{\bf k}}}
\newcommand{\hs}{{\hat s}}

\newcommand{\hsigma}{{\hat\sigma}}

\newcommand{\cD}{{\cal D}}

\newcommand{\cH}{\hat{\cal H}}
\newcommand{\cL}{{\cal L}}

\newcommand{\cO}{{\cal O}}

\def\holOne{\mathds{1}}

\newcommand{\dis}{\text{dis}}
\newcommand{\HP}{\text{Hop}}
\newcommand{\SM}{\text{SM}}

\newcommand{\Average}[1]{\left< #1 \right>}

\begin{document}
\title{Duality between disordered nodal semimetals and systems with power-law hopping}

\author{S.V.~Syzranov}
\affiliation{Physics Department, University of California, Santa Cruz, California 95064, USA}

\author{V.~Gurarie}
\affiliation{Department of Physics and Centre for Theory of Quantum Matter,
	University of Colorado, Boulder, Colorado 80309, USA}

\begin{abstract}
Nodal semimetals (e.g. Dirac, Weyl and nodal-line semimetals, graphene, etc.) and systems 
of pinned particles with power-law interactions (trapped ultracold ions, nitrogen defects in diamonds, spins
in solids, etc.) are presently at the
centre of attention of large communities of researchers working in condensed-matter and atomic, molecular
and optical physics. Although seemingly unrelated, both classes of systems are abundant with novel 
fundamental thermodynamic and transport phenomena.
In this paper, we demonstrate that low-energy field theories of quasiparticles in semimetals
may be mapped exactly onto those of pinned particles with power-law-hopping excitations.
The duality between the two classes of systems, which we establish, trades strong disorder in one class for weak disorder in the other, and allows one to describe the transport 
and thermodynamics of each class of systems using the results established for the other class.
In particular, using the duality mapping, we establish the existence of a novel class of disorder-driven transition in systems with the power-law hopping $\propto1/r^\gamma$ of excitations
with $d/2<\gamma<d$, different from the conventional Anderson-localisation transition. Non-Anderson disorder-driven
transitions have been studied broadly for nodal semimetals, but have been unknown, to our knowledge, for 
systems with long-range hopping (interactions) with $\gamma<d$.
\end{abstract}


\maketitle


The last few years have seen an explosion of interest in nodal semimetals, such
as Weyl, parabolic and nodal-line semimetals~\cite{Armitage:WeylReview}, owing to their potential applications in future electronic and spintronic devices, in addition to the abundance of novel 
fundamental phenomena observed in these materials: chiral anomaly~\cite{Burkov:review,Parameswaran:ChiralAnomaly,Huang:anomalyObservation,Ong:anomalyExperiment,Burkov:diffusiveAnomaly}, magnetohydrodynamic effects~\cite{Gorbar:WeylHydro,LucasSachdev:WeylHydrodynamics,Yamamoto:HEPhydrodynamics,Galitski:dynamo}, topologically protected surface states~\cite{WanVishwanath:WeylFirst,Weng:drumheadStates,Hasan:reviewWeyl},
etc.
These systems have also 
changed the current perspective on phase transitions in disordered systems;
 Weyl and Dirac semimetals have been demonstrated to display 
 disorder-driven phase transitions (or possibly very sharp
crossovers~\footnote{\label{foot1} It is still being debated in the literature (see, e.g., 
	Refs.~\onlinecite{Nandkishore:rare,Sbierski:WeylVector,Mafia:rare,Gurarie:AvoidedCriticality,BuchholdAltland:rare,BuchholdAltland:rare2}) whether the
	respective transition in a Weyl semimetal is a genuine phase
	transitions or a sharp crossover in physical observables. In this paper we do not distinguish between transitions and sharp crossovers, so long as there exists a parametrically large interval of observables 
	where the critical scaling is observed.}) in universality classes  
different from those of the Anderson metal-insulator transitions~\cite{Fradkin1,Fradkin2,Syzranov:review}.
These transitions have been demonstrated~\cite{Syzranov:WeylTransition} to occur also in a broader class of systems in sufficiently high
dimensions.

Another, seemingly unrelated, class of systems, which have recently attracted attention, are systems of pinned particles with power-law interactions. 
The amplitude of hopping of excitations in these systems displays
a power-lay decay with distance, $\propto1/r^\gamma$.
Such systems include, but are not limited to, polar molecules~\cite{dipolar_experiment_2013,dipolar_review_2009} ($\gamma=3$),
impurity spins in solids, Rydberg atoms~\cite{Rydberg_Review_2010} ($\gamma=3$ or $\gamma=6$), nitrogen vacancies in diamonds~\cite{Dutt:diamonds,Waldherr:diamonds} ($\gamma=3$) and neutral excitations in strongly
disordered electronic systems ($\gamma=3$ in 3D systems and $\gamma=2$ in thin dielectric films~\cite{Mooij:electrostatics,AleinerEfetov:dipoles,Titum:LevelStatistics}).
Furthermore,
power-law hopping with a tunable parameter $0<\gamma<3$ has been realised~\cite{Monroe:longrange,Islam:longrange,Blatt:chain1,Blatt:chain2} recently in 1D and 2D systems of trapped ultracold ions. 
Similarly to the case of nodal semimetals, in the presence of quenched disorder
power-law hopping is expected to lead to unconventional
localisation phenomena or disorder-driven criticality (see, e.g., Refs.~\cite{Levitov2,Levitov:AnnReview,Levitov_1990,Deng:powerLawDuality,
	Rodriguez:firstPowerLaw,Malyshev:firstPowerLaw,Moura:firstPowerLaw,Garttner:longrange,TikhnonovMirlin:AlphaSimulations,Yao:Levitov}) and is also studied often in the context 
of many-body localisation~\cite{BAA,NandkishoreHuse:review,Abanin:PowerLawReview,AbaninAltman:MBLreview,Burin:LevitovsRG,Burin:MBLclaims}.

Nodal semimetals and systems with power-law hopping may seem completely unrelated 
and are usually studied independently by two communities of researchers working, respectively,
on condensed-matter systems and in atomic, molecular and optical physics. 
As it has already been noted~\cite{Garttner:longrange,Syzranov:review}, however, 
$d$-dimensional systems with power-law hopping, whose amplitude decays with distance as $\propto 1/r^\gamma$ with $d<\gamma<3d/2$,
display the same type of non-Anderson disorder-driven transitions which take place in nodal semimetals
and which are absent for faster decay, corresponding to $\gamma>3d/2$.
There is thus a natural connection between the two classes of systems.
The phenomenology of systems with slower decay of the hopping, corresponding to $0<\gamma<d$, is understood,
in our opinion, much more poorly; the possibility of the unconventional 
disorder-driven transitions in them has not been investigated until this work. 

\begin{figure*}[ht!]
	\centering
	\includegraphics[width=0.9\textwidth]{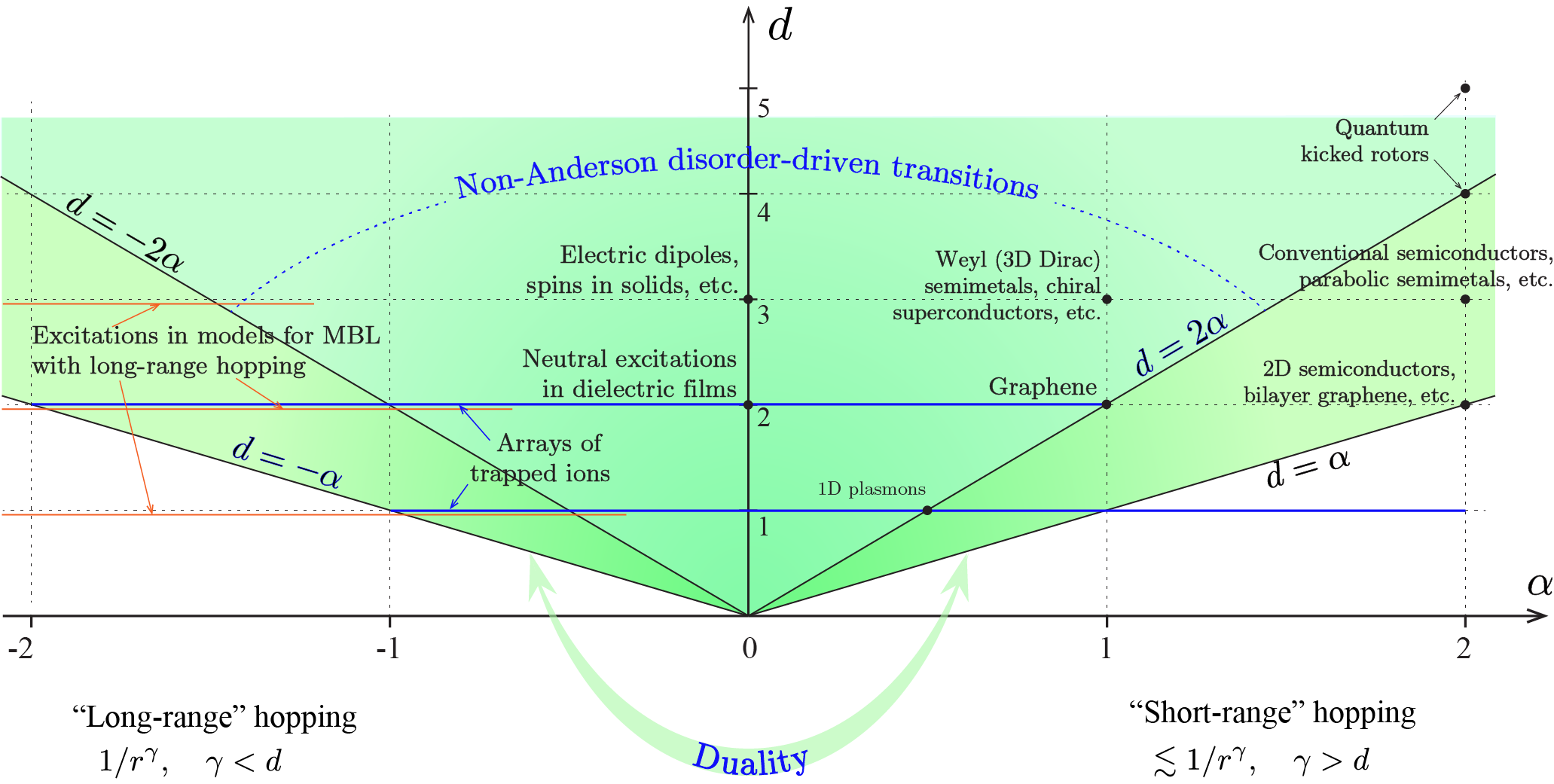}
	\caption{\label{fig:dualitydiagram}
		(Colour online) Examples of systems with the power-law dispersion $\xi_\bk\propto k^\alpha$ in the ``dimension $d$ vs. $\alpha$ diagram''.
		$\alpha>0$ corresponds to nodal semimetals 
		and systems where hopping decays with distance faster than $1/r^d$.
		Systems with $\alpha<0$ are represented
		by trapped ultracold particles with long-range interactions and models of many-body-localisation (MBL) transitions with power-law hopping. The duality transformation, described in this paper, maps the region $d>\alpha>0$ of the diagram onto the region $d>-\alpha>0$. This duality allows us to predict previously unknown disorder-driven phase transition in systems with long-range hopping, 
		which correspond to $d>-2\alpha>0$.}		
\end{figure*}

In this paper, we 
derive a duality transformation between the field theories of disordered systems with
the power-law dispersion $\propto k^\alpha$, with $0<\alpha<d$, typical for nodal semimetals,
and the field theories of hopping in arrays of random-energy sites with the amplitude
$\propto 1/r^\gamma$, with $0<\gamma<d$, hereinafter referred to as 
{\it ``long-range hopping''} (as opposed to {\it ``short-range hopping''}, decaying with distance 
faster than $1/r^d$). Since particles displaying the long-range hopping 
between sites may be considered as 
having a singular kinetic energy
$\propto k^{\gamma-d}$, 
the duality transformation may be said to
establish a mapping between disordered systems with kinetic energies $k^\alpha$ and $k^{-\alpha}$,
as shown in Fig.~\ref{fig:dualitydiagram}, where $0<\alpha=\gamma-d<d$. 
It is of note that this duality exchanges strong and weak disorder in the two phases it connects.
This duality mapping may be used to
describe phenomena in each of these two classes of systems,
by using theoretical descriptions and experimental results available for the other class, or to
develop new models and descriptions.

Using the established duality, we predict a novel class of disorder-driven quantum phase transitions 
in systems with long-range hopping $\propto 1/r^\gamma$ with $d/2<\gamma<d$.
These transitions are dual to   
the non-Anderson disorder-driven transitions~\cite{Syzranov:review} in 
nodal semimetals and systems with short-range hopping (see Fig.~\ref{fig:dualitydiagram}).
While unconventional disorder-driven transitions have been a target
of vigorous studies in the context of Weyl semimetals (see Ref.~\cite{Syzranov:review} for a review)
and of short-range hopping~\cite{Rodriguez:firstPowerLaw,Malyshev:firstPowerLaw,Moura:firstPowerLaw,Garttner:longrange,Syzranov:multifract} (i.e. decaying faster than $1/r^d$),
disorder-driven transitions for long-range hopping $\propto1/r^\gamma$, with $\gamma<d$,
have not been investigated previously, to the best of our knowledge, and are established here for the
first time by means of the duality arguments.

These transitions manifest themselves in the singular behaviour of the density of states (DoS)
and other observables. We emphasise that these transitions are not accompanied by 
localisation or delocalisation of wavefunctions, as all states are expected to be delocalised
for long-range hopping~\cite{Levitov2,Levitov_1990}, and belongs to a universality class different from 
that~\cite{EversMirlin:review,Efetov:book} of Anderson localisation. 
These transitions may be realised by means of ultracold ions in optical or magnetic
traps~\cite{Monroe:longrange,Islam:longrange,Blatt:chain1,Blatt:chain2} and also occur for excitations 
in certain models which are used usually for studying many-body localisation-delocalisation transitions in the presence of power-law
hopping~\cite{Burin:LevitovsRG,Burin:MBLclaims,GutmanMirlin:powerLaw}.

%

{\it Duality mapping.}
To establish the duality, we consider 
a semimetal with the power-law quasiparticle dispersion $\xi=ak^\alpha$ near the node,
which has a trivial spin and sublattice structure, 
in the presence of randomly located short-range impurities, whose potentials may be approximated
by delta-functions with amplitudes $\lambda_n$; $u(\br)=\sum_n\lambda_n\delta(\br-\br_n)$. Our arguments
may easily be generalised to the cases of more complicated dispersions, including spin and valley structures,
and impurity potentials~\cite{Supplemental}. In this paper, we focus on single-particle disorder-driven phenomena
and use a supersymmetric~\cite{Efetov:book} field theory to describe quasiparticle states whose action is given by 
\begin{align}
	\cL_\SM=-i\int \bar\psi\left[E-a|\hbk|^\alpha+i0\Lambda\right]\psi\, d\br
	\nonumber\\
	+i\sum_n \lambda_n \bar\psi(\br_n)\psi(\br_n),
	\label{ActionSemiMetal}
\end{align}
where 
$\hbk=-i\partial_\br$ is the momentum operator;
$\psi$ is a four-component supervector with components in the $BF\otimes RA$ (boson-fermion $\otimes$
retarded-advanced) space~\cite{Efetov:book}; $\Lambda=\holOne_{BF}\otimes\left(\hsigma_z\right)_{RA}$,
and $\bar\psi=\psi^\dagger\Lambda$.

Values of observables in this semimetal may be represented in the form $\Average{\cO}=\int D\psi D\psi^\dagger \ldots e^{-\cL_\SM}$, where $\ldots$ are supersymmetry-breaking terms.
Introducing supervectors $\phi(\br_n)$ at the locations $\br_n$ of the impurities by means
of the Habbard-Stratonovich transformation
$e^{-i\lambda\bar{\psi}\psi}=\int \cD\phi^\dagger\cD\phi 
e^{i\frac{\bar\phi\phi-\lambda\bar\psi\phi-\lambda\bar\phi\psi}{\lambda-i0\Lambda}}$ 
and integrating out the fields $\bar\psi$ and $\psi$ leads
to a field theory with the action
\begin{align}
	\cL=i\sum_{m,n}\bar{\phi}(\br_n)G_E(\br_n,\br_m)\phi(\br_m)
	-i\sum_n\frac{\bar{\phi}(\br_n)\phi(\br_n)}{\lambda_n-i0\Lambda},
\end{align}
in terms of the fields $\phi$, where 
$G_E=\int\frac{d\bk}{(2\pi)^d}\frac{e^{-i\bk\br}}{E-a |\bk|^\alpha+i0\Lambda}$ is a matrix of the 
Green's functions in a disorder-free semimetal.

In the vast majority of nodal semimetals (such as 3D Dirac, Weyl and parabolic semimetals, graphene) the exponent $\alpha$
of the dispersion $\xi_\bk\propto |\bk|^\alpha$
is exceeded by the dimensions $d$. Power-law dispersion with $\alpha<d$ may also be realised in 1D and 2D systems of trapped
 ultracold ions~\cite{Monroe:longrange,Islam:longrange,Blatt:chain1,Blatt:chain2} and certain other systems, such as 1D
 plasmons (corresponding to $\alpha=1/2$).  
In all these systems, 
the Green's functions $G_E(\br)$ at low energies $E$ display power-law decay with distance $r$, which 
gives the action
\begin{align}
&\cL_\HP=
\nonumber\\
&-i\sum_{m,n}\bar{\phi}(\br_n)\left[\frac{j}{|\br_n-\br_m|^{d-\alpha}}
+\delta_{nm}\lambda_n^{-1}
+i0\Lambda\right]\phi(\br_m),
\label{ActionHopping}
\end{align}
where $j=\frac{\Gamma\left(\frac{d-\alpha}{2}\right)}{2^\alpha a \pi^\frac{d}{2}\Gamma\left(\frac{\alpha}{2}\right)}$
for the case of the dispersion $\xi_\bk=a|\bk|^\alpha$ with a trivial spin and valley structure considered here.
For more complicated structures of the dispersion of the semimetal, the first term in the action \eqref{ActionHopping} will
still display the power-law decay $\propto 1/|\br_n-\br_m|^{d-\alpha}$, but will involve the respective valley or spin degrees of freedom. 
For example, in the case of a Weyl semimetal, with the quasiparticle dispersion given by $\xi_\bk=v\hbsigma\cdot \bk$, where $\hbsigma$ is the pseudospin operator,
the system may be described by action \eqref{ActionHopping} with the replacement
$j\rightarrow -\frac{1}{8\pi}\frac{\hbsigma\cdot(\br_n-\br_m)}{v|\br_n-\br_m|^3}$.

Action~\eqref{ActionHopping} describes particle hopping between randomly located sites, with the energies
$-\lambda_n^{-1}$, 
where the hopping amplitude decays with distance as the power-law $\propto1/|\br_n-\br_m|^{d-\alpha}$.
The transformation from action~\eqref{ActionSemiMetal} to action~\eqref{ActionHopping}
is, therefore, a duality transformation between
the field theories of a disordered nodal semimetal and a systems with power-law hopping in an array of 
random-energy sites.

This duality may be used to explore novel phenomena in each of these two classes of systems
using the theories and experimental results available for the other class. 
In particular, we predict in what follows a novel disorder-driven 
quantum phase transition for systems with power-law hopping, which is dual to
the disorder-driven transition in semimetals in high dimensions~\cite{Syzranov:review}.

{\it Non-Anderson disorder-driven transitions in systems with long-range hopping.} In order to elucidate these novel transitions, we consider a spin-$1/2$ $XY$ model on randomly located sites, where the exchange interaction
between the spins depends on the distance between them as the power law. The Hamiltonian of this model, which may be
realised by means of trapped ultracold particles, is given
by 
\begin{align}
\cH=\sum_n E_n\hs_n^z-\sum_{n,m} \frac{j}{|\br_{n}-\br_m|^\gamma}\left(\hs_n^+\hs_m^-+\hs_m^+\hs_n^-\right),
\label{SpinModel}
\end{align}
where the on-site magnetic fields $-E_n$ are random and uncorrelated on different sites
and we consider $\gamma<d$.

In general, in the limit of small $j$, spin-flip-type excitations in this model propagate similarly to single particles
with the amplitudes which decay with distance as the power law~\cite{FleishmanAnderson,AleinerEfetov:dipoles,Titum:LevelStatistics,GutmanMirlin:powerLaw,Yao:Levitov}.
For simplicity, we assume that the on-site field $E_n$ fluctuates weakly on top of a large average value $\Average{E_n}_\dis$
 and, as a result,
all the spins are almost entirely polarised in the same direction in the ground state.
The propagation of spin-flip excitations may then be mapped exactly onto a single-particle model
with the action~\eqref{ActionHopping}. 

In what follows,
we demonstrate that the system exhibits a non-Anderson phase transition when changing 
the coupling amplitude $j$ or the amplitude of the fluctuations of the on-site energies $E_n$.
This transition may manifest itself in the form of singularities in various physical observables,
such as the diffusion coefficient or magnetic susceptibility. In what follows we focus on the 
behaviour of the DoS of low-energy excitations.

Below, we consider the states of the Hamiltonian~\eqref{SpinModel} with energies $E=\Average{E_n}_\dis+\omega$ weakly deviating from the large average on-site 
field $\Average{E_n}_\dis$.
The DoS of single spin-flip excitations at energy $\omega$ 
is given by
\begin{subequations}
\begin{align}
	\rho_\HP(\omega) &=-\frac{1}{\pi V}\text{Im}\, 
	\partial_\eta\Average{\int\cD\phi\cD\phi^\dagger
	e^{-{\cL_\HP-\cL_\omega}-\cL_s(\eta)}}_\dis,
	\label{DoSstarting}
	\\
	\cL_\omega &=-i\omega\sum_n \bar{\phi}(\br_n){\phi}(\br_n),
	\label{OmegaTermStarting}
	\\
	\cL_s(\eta) &=i\eta\sum_n s_R^*(\br_n) s_R(\br_n),
	\label{SourceTermStarting}
\end{align}
\end{subequations}
where $s_R(\br_n)$ is the retarded bosonic component of the supervector $\phi(\br_n)$;
$V$ is the volume of the system; $\cL_\HP$ is given by Eq.~\eqref{ActionHopping} with 
$\lambda_n^{-1}=- E_n+\Average{E_n}_\dis$;
the term $\cL_\omega$ accounts for the effect of the energy $\omega$ on the action; $\cL_s(\eta)$
is the supersymmetry-breaking source term; $\eta$
is an infinitesimal parameter.

The full action $\cL=\cL_\HP+\cL_\omega+\cL_s(\eta)$ used to obtain the DoS \eqref{DoSstarting}
thus matches action \eqref{ActionHopping} with the replacement 
$\lambda_n^{-1}\rightarrow -E_n+\Average{E_n}_\dis+\omega-\eta\cdot\holOne_{BBRR}$, where $\holOne_{BBRR}$ is the 
projector
to the bosonic retarded parts of supervectors. 
The corresponding dual action of a semimetal is given by Eq.~\eqref{ActionSemiMetal} with the
same replacement and with $\alpha=d-\gamma$ and $a=
\frac{\Gamma\left(\frac{\gamma}{2}\right)}
{2^{d-\gamma}\pi^\frac{d}{2}\Gamma\left(\frac{d-\gamma}{2}\right)j}$.

Expanding that action to the first order in small parameters $\omega$ and $\eta$ and performing
disorder averaging of the DoS, while keeping only the first cumulants in the disorder
strength
\begin{align}
	\varkappa=N\left<\left(E_n-\Average{E_n}\right)^{-2}\right>_\dis,
	\label{DisorderStrengthRaw}
\end{align}
gives the effective action of a semimetal
\begin{align}
	\cL_\SM=&-i\int \bar\psi\left[\omega \varkappa-a|\hbk|^{d-\gamma}+i0\Lambda\right]\psi\, d\br
	\nonumber\\
	&+\frac{1}{2}\varkappa\int \left[\bar{\psi}\psi\right]^2 d\br
	+i\eta \varkappa \int \tilde s_R^*\tilde s_R d\br,
\label{ActionSemiMetalFinal}
\end{align}
where $\tilde s_R^*$ and $\tilde s_R$ are the retarded bosonic components of the supervectors $\bar{\psi}$
and $\psi$.
We emphasise that we assume, when deriving Eq.~\eqref{ActionSemiMetalFinal},
that the variance $\varkappa^2/N$ of the quantity $\left(E_n- \left< E_n \right> \right)^{-1}$ 
is finite. For example, this quantity may be a random Gaussian variable, in which case
the fluctuations of the energies $E_n$ are non-Gaussian; the statistics or even the existence of the 
variance of the energies $E_n$ have no bearing on our arguments.

Equation \eqref{ActionSemiMetalFinal} is the action of a disordered semimetal with the 
quasiparticle dispersion $\xi_\bk=a k^{d-\gamma}$ and with a source term which generates the DoS. According to Eq.~\eqref{DoSstarting} and the action \eqref{ActionSemiMetalFinal}, the DoS of low-energy excitations in the model described by the Hamiltonian~\eqref{SpinModel}
is given by 
\begin{align}
\rho_\HP(\omega)=\varkappa \rho_\SM (\omega\varkappa),
\label{DoS}
\end{align}
where $\rho_\SM(\omega\varkappa)$ is the DoS in the dual semimetal described 
by the action~\eqref{ActionSemiMetalFinal} at energy $\omega\varkappa$ and the disorder strength $\varkappa$
given by Eq.~\eqref{DisorderStrengthRaw}.

Semimetals described by the action of the form~\eqref{ActionSemiMetalFinal}
(Dirac semimetals, high-dimensional semiconductors, quantum kicked rotors, etc.) are known 
to display phase transitions (or sharp 
crossovers~\cite{Note1}) between strong-disorder and weak-disorder phases
at a certain value $\omega_c$ of the energy, hereinafter set to zero, and disorder strength. 
\begin{figure}[t]
	\centering
	\includegraphics[width=\linewidth]{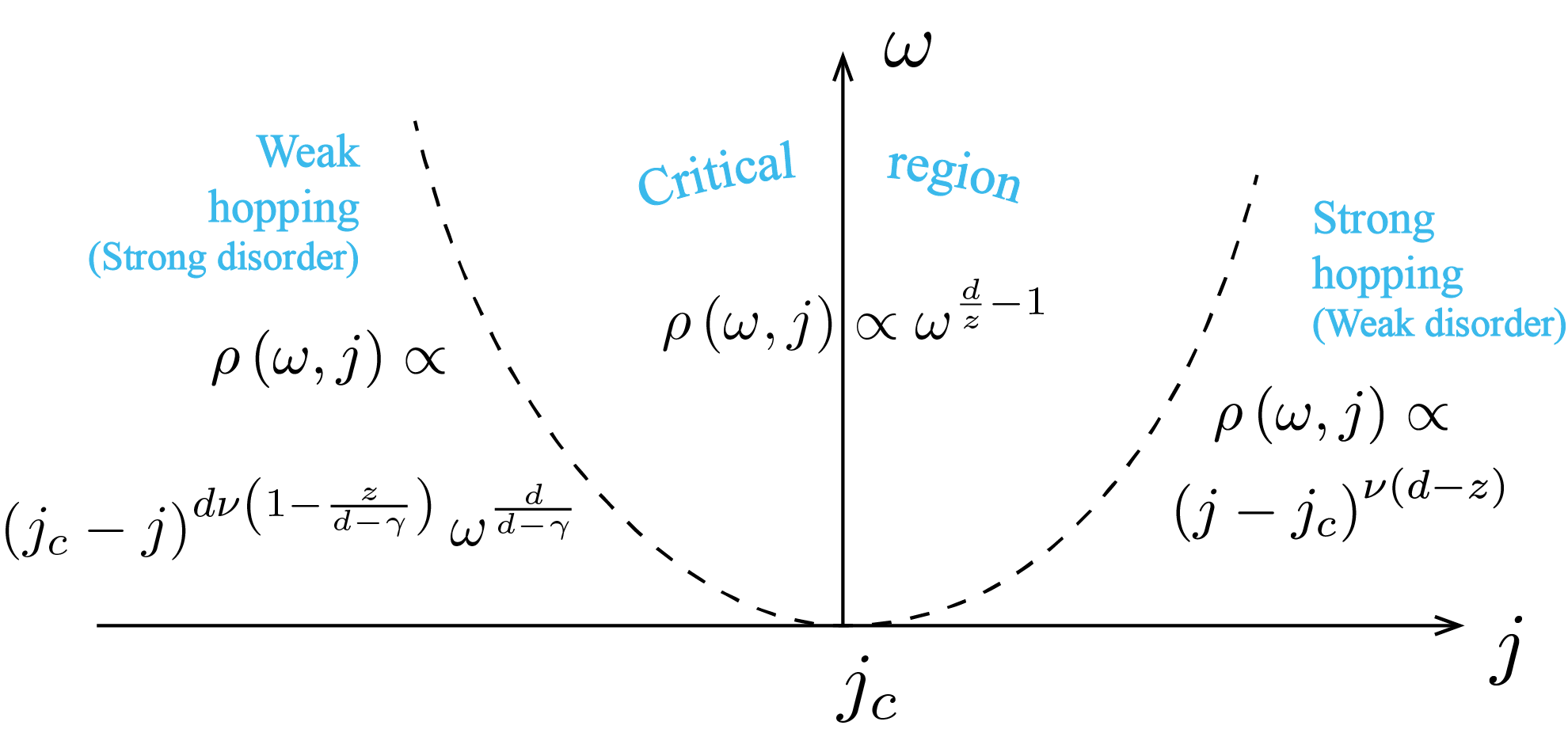}
	\caption{	\label{fig:phasediagram} (Colour online) Critical behaviour of the density of states
	of low-energy excitations in a system with the power-law hopping $\propto j/r^\gamma$
	in the diagram ``excitation energy $\omega$ vs. coupling $j$'', where the energy $\omega$ is measured from the average potential.
	The same diagram with the replacement $j\rightarrow 1/\varkappa$ describes the dependence
	of the DoS on energy $\omega$ and the disorder strength $\varkappa$.   
	}
\end{figure}
The relation \eqref{DoS} between the densities of states of a semimetal and a model with
power-law hopping indicates that a disordered system with long-range hopping exhibits a critical 
scaling in the same universality class.
However, because the on-site energies $E_n$ in the hopping model 
match the inverse amplitudes $\lambda_n^{-1}$ of the impurity potentials in 
the dual semimetal, the strong- and weak-disorder phases of the semimetal corresponds,
respectively, to the weak- and strong-disorder phases of system with long-range hopping.
If the disorder strength $\varkappa$ exceeds a critical value $\varkappa_c$ in a system with long-range 
hopping,
or the coupling $j$ is smaller than a critical coupling $j_c$, the system
is in
a phase with a suppressed DoS of low-energy excitations. 
At weaker disorder or larger intersite couplings, the systems exhibits a continuous transition~\cite{Note1} to a phase with a larger DoS and better transport of the low-energy excitations.

Using Eq.~\eqref{DoS} and the critical scaling proposed in Ref.~\cite{KobayashiOhtsukiHerbut:scaling}
for a Dirac semimetal
and derived microscopically in Ref.~\cite{Syzranov:unconv} for a generic semimetal with
the power-law quasiparticle dispersion $\xi_\bk\propto k^\alpha$, we find that 
the DoS near the transition in a
system with long-range hopping has the form 
\begin{align}
	\rho_\HP(\omega,j)=\omega^{\frac{d}{z}-1}\Phi\left(\omega/|j-j_c|^{z\nu}\right)
	+\rho_{\text{smooth}},
	\label{scaling}
\end{align}
where $\Phi(x)$ is a universal scaling function which may be different for different signs of the argument,
and $\nu$ and $z$ are the correlation-length and the dynamical critical exponents (matching those of the dual transition
in a semimetal~\cite{GoswamiChakravarty,PixleyHuse:missedPoint,Sbierski:WeylVector,LiuOhtsuki:LateNumerics,LouverCarpentier:longRange,Malyshev:firstPowerLaw,Syzranov:review}); $\rho_{\text{smooth}}$
is an analytic contribution which comes from the instantons in actions
\eqref{ActionSemiMetal} and \eqref{ActionHopping} 
(``rare-region effects''~\cite{Wegner:DoS,Suslov:rare,Nandkishore:rare,Syzranov:unconv,Mafia:rare,Gurarie:AvoidedCriticality,BuchholdAltland:rare,BuchholdAltland:rare2}). 

In semimetals, the instantonic contribution $\rho_{\text{smooth}}$ to the DoS
may be exponentially suppressed by the small deviation $|d-d_c|$ of the dimension $d$ from the 
critical dimension $d_c=2\gamma$
 of the transition or the number of the particle flavours~\cite{Syzranov:review}.
Even in the absence of small parameters, various numerical studies of 3D Weyl and Dirac semimetals 
have found this contribution to be rather small or 
unobservable~\cite{KobayashiOhtsukiHerbut:scaling,Sbierski:firstWeyl,Pixley:twoTransitions,BeraRoy:numerics,LiuOhtsuki:LateNumerics,PixleyHuse:missedPoint,Sbierski:WeylVector,Wang:multifractality,Syzranov:review}, which allows one to use the DoS, to a good approximation, as an order parameter for the transition. It has also been suggested~\cite{Note1} that
the instantonic contribution may broaden criticality, thus converting the transition to a sharp
crossover. In this paper, however, we do not distinguish between such sharp crossovers and 
phase transitions.
The behaviours of the DoS for various values of the coupling $j$ and the excitation energies $\omega$ near the critical point 
are summarised in Fig.~\ref{fig:phasediagram} and
follow directly from the scaling form \eqref{scaling} and may also be inferred from the respective results 
for semimetals~\cite{KobayashiOhtsukiHerbut:scaling,Syzranov:review}. 

{\it Experimental observation.} The DoS of excitations
in a spin model with the Hamiltonian~\ref{SpinModel} may
be observed explicitly in experiments on trapped ultracold particles by measuring the spin susceptibility
$\chi_{xx}(\omega)=i\sum_j\int_0^\infty \left<\left[\hs_i^x(t),\hs_j^x(0)\right]\right>e^{it(\omega+i0)}dt$.
The imaginary part of the susceptibility determines the dissipation in the system due to creating 
spin-flip excitations and is related to the density of states of the excitations as  
$\text{Im}\chi_{xx}(\omega)=N\rho(\omega)$.

Describing spin-flip excitations in the spin model with the Hamiltonian~\eqref{SpinModel} by a single-particle model
with action~\eqref{ActionHopping}, considered here, is justified
if all the spins are polarised in the ground state by the average value of the on-site magnetic magnetic field.
It is also expected often that single-particle descriptions for excitations in the considered spin model are appropriate 
even if the sign of the magnetic field fluctuates in space, so long as the parameter $j$ is sufficiently small~\cite{FleishmanAnderson,AleinerEfetov:dipoles,Titum:LevelStatistics,GutmanMirlin:powerLaw,Yao:Levitov}.
In the latter case, we still expect the 
existence of disorder-driven phase transitions between the ``weak-hopping'' (irrelevant ratio of the hopping 
to disorder amplitudes) phase at small $j$
and a ``strong hopping'' phase at larger $j$. We leave, however, such
models for future studies.

{\it Outlook.}
Natural further research directions include extending the duality between nodal semimetals and systems
with long-range hopping, which we have established here, to interacting systems, more generic and anisotropic
dispersions (e.g., in nodal-line and nodal-surface semimetals) and other models of disorder.
The new approach of describing semimetals in the dual representation, established here, may also 
also be used to investigate, for example, rare-region effects and the possibility of other unconventional 
phase transitions.

{\it Acknowledgements.} We are grateful to L.~Radzihovsky and B.~Sbeirski for useful discussions
and feedback on the manuscript.
Also, we thank L.~Radzihovsky for prior collaboration
on related topics. 
Our work has also been supported by the Hellman Foundation (SVS) and the 
Faculty Research Grant awarded by the Committee on Research from the University of California,
Santa Cruz (SVS).


%


\onecolumngrid
\vspace{2cm}

\cleardoublepage

\renewcommand{\theequation}{S\arabic{equation}}
\renewcommand{\thefigure}{S\arabic{figure}}
\renewcommand{\thetable}{S\arabic{table}}
\renewcommand{\thetable}{S\arabic{table}}
\renewcommand{\bibnumfmt}[1]{[S#1]}
\renewcommand{\citenumfont}[1]{S#1}

\setcounter{equation}{0}
\setcounter{figure}{0}
\setcounter{enumiv}{0}

\pagestyle{empty}

\begin{center}
	\textbf{\large Supplemental Material for \\
		``Duality between disordered nodal semimetals and systems with power-law hopping''
	}
	\\
	S.V.~Syzranov and V.~Gurarie
\end{center}
\vspace{5mm}

\section{Continuous vs. discrete duality transformations}

In the main text we focussed, when deriving the duality transformation,
on quasiparticles scattered off randomly located $\delta$-impurities and mapped it onto
a model of hopping between discrete sites at the locations of the impurities.
This duality transformation may be generalised straightforwardly to the case of a
random potential $u(\br)$ which varies in space continuously.

The value of an observable $\cO$ in a semimetal with the kinetic
energy $\xi_\bk$ with a continuous random potential is given by
\begin{align}
\Average{\cO}=\int D\psi D\psi^\dagger \ldots \exp\left[{i\int \bar\psi\Lambda\left[E-\xi_{\hat k}+i0\Lambda\right]\psi d\br-
	i\int u(\br) \bar\psi(\br)\psi(\br)}d\br\right],
\label{StartingActionCont}
\end{align}
where $\ldots$ are the supersymmetry-breaking terms corresponding to this observable.
The Hubbard-Stratonovich transformation
\begin{align}
\int D\phi D\phi^\dagger \exp\left[{i\int\frac{(\bar\phi-u \bar\psi)(\phi-u \psi)}{u-i0\Lambda}}d\br\right]=1,
\end{align}
leads to the dual representation
\begin{align}
\Average{\cO}=\int D\phi D\phi^\dagger
\ldots \exp\left[-i\int\bar{\phi}(\br)G(\br,\br^\prime)\phi(\br^\prime)\,d\br d\br^\prime
+i\int\frac{1}{u(\br)-i0\Lambda}{\bar{\phi}(\br)\phi(\br)}d\br\right],
\label{DualActionCont}
\end{align}
where for a power-law dispersion 
$\xi_\bk\sim a k^\alpha$
the Green's function $G(\br,\br^\prime)$ displays a power-law decay
$\propto 1/|\br-\br^\prime|^{d+\alpha}$ for low energies $E$.

Equations \eqref{StartingActionCont} and \eqref{DualActionCont} represent a duality mapping
between a nodal semimetal in a continuous potential $u(\br)$ and a systems with power-lay hopping
in the potential $1/u(\br)$. 
In principle, models with continuous random potentials may in general be approximated, respectively, by
models with discrete short-ranged impurities and models on discrete sites, considered in the 
main text, by coarse-graining the potential at sufficiently small length scales.

%
%
%
%
%
%

\end{document}